\documentstyle[aps,version2,preprint,epsfig]{revtex}
\begin{document}
\draft
\preprint{OCIP/C 97-04}
\preprint{April 1997}
\begin{title}
Comment on $Z'$'s and the H1 and ZEUS High $Q^2$ Anomalies 
\end{title}
\author{Stephen Godfrey}
\begin{instit}
Ottawa-Carleton Institute for Physics \\
Department of Physics, Carleton University, Ottawa CANADA, K1S 5B6
\end{instit}

\begin{abstract}
We investigate the effects of extra neutral gauge bosons on the high 
$Q^2$ region of the $e^+p \to e^+ X$ cross section at 
$\sqrt{s}=300$~GeV.  We found that the only models with electroweak 
strength coupling, typical of extended gauge theories, that 
give a better fit to the H1 and ZEUS high $Q^2$ data than
the standard model,
are ruled out by existing data from the Tevatron.  From general 
scaling arguments, using the allowed contact interactions, the only 
allowed models with $Z'$'s would be those with strong couplings
although even in this case the statistical evidence is not compelling.  

\end{abstract}

The H1 \cite{h197} and ZEUS \cite{zeus97} experiments at HERA have 
observed an excess of events in $e^+ p \to e^+ X$ at $Q^2 > 
15,000$~GeV$^2$ compared to Standard Model expectations in 
34.3~pb$^{-1}$ of accumulated luminosity.
These  observations have led to considerable speculation
\cite{lq,squarks,ci,ci2,misc}.  Although these 
observations cannot be ruled out to be statistical fluctuations the 
two collaborations estimate that this is unlikely.  Some of the more 
popular interpretations of the high $Q^2$ excess are that it signals the 
existence of leptoquarks \cite{lq}, squarks \cite{squarks}, or contact 
interactions of quarks and leptons \cite{ci,ci2}.  In the latter case, a 
number of authors have suggested that the contact interaction 
represents a low energy effective interaction which parametrizes 
the t-channel exchange of an extra neutral gauge boson ($Z'$) 
\cite{ci,ci2,wm}.
In this note we explore the viability of this explanation of the HERA 
high $Q^2$ anomaly.

Extra neutral gauge bosons are a feature of many extensions of the 
standard model such as grand unified theories, left-right symmetric 
models, excited weak vector bosons, and a strong Higgs sector
\cite{c-g,godfrey,tom,er5m,hyp,baur,bess,othermodels,layssac,bess2}.  
$Z'$'s contribute to $ep$ cross sections via t-channel exchange and the 
resulting interference with standard model contributions \cite{capstick}.  
The expressions for the resulting cross sections have been given elsewhere 
so for brevity we do not reproduce them here but refer the interested 
reader to previously published results \cite{capstick}.  

To quantify the effect of a $Z'$ on the HERA cross sections we compare 
the cross section with a $Z'$ to the combined H1 and ZEUS data 
combined bin-by-bin for bins starting at $Q^2=(5000, \; 10000, \; 15000, \; 
20000, \; 25000, \; 30000, \; 35000)$~GeV$^2$ with each bin 
being 5000~GeV$^2$ wide except the last which goes from 
$Q^2=35000$~GeV$^2$ to the kinematic limit. The calculated results 
were corrected for detection efficiencies of 80\% and include an 
average QCD K-factor of 1.1.  We use a log-likelihood procedure
with Poisson 
statistics, appropriate for small numbers of events, to fit $M_{Z'}$ 
for a large number $Z'$'s arising from 
models with extended gauge groups\cite{tom,er5m,hyp,othermodels}.  

In only two models did including a $Z'$ 
improve the fit and in no case would we argue that the 
improved fit was statistically distinguishable from the standard 
model\footnote{We note that using a $\chi^2$ analysis results in a 
lower $M_{Z'}$ and implies a better fit than the 
log-likelihood approach.  However the log-likelihood approach is more 
appropriate for low statistics.}.
In Fig. 1 we plot the $e^+p$ differential cross 
sections as a function of the minimal $Q^2$ value
for the standard model and two cases, $M_{Z_{\psi}}=150$~GeV \cite{er5m}
\footnote{Varying $\theta_{E_6}$, a parameter of $E_6$ GUTS 
\cite{capstick}, we obtain 
the best fit for $\theta_{E_6}=\pi/2$ which correponds to $Z_\psi$.}
and $M_{Z_{HYP}}=420$~GeV \cite{hyp}
along with the HERA results which are corrected for 
detector efficiencies of 80\% and divided by an average QCD K-factor 
of 1.1  so that we show leading order cross sections.  
 
\vskip 0.4cm
\centerline{\epsfig{file=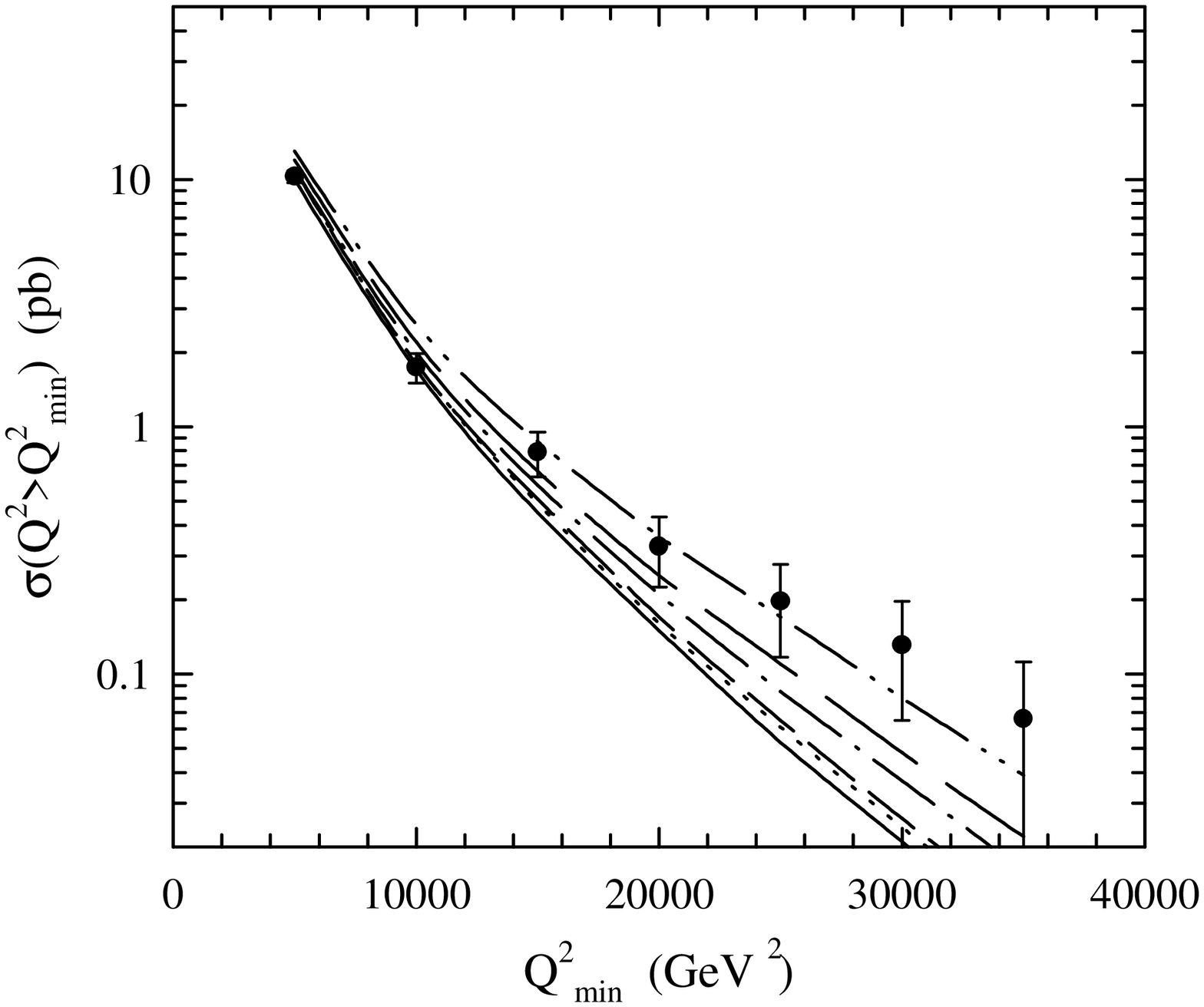,width=8.5cm,clip=}}
\noindent
{\bf Fig 1:} Integrated cross sections versus a minimum $Q^2$ for $e^+p \to 
e^+ X$ for the SM (solid curve), $M_{Z_{\psi}}=150$~GeV (dotted curve),
$M_{Z_{HYP}}=420$~GeV (medium-dashed curve) and the contact 
interactions with $\eta_{LR}^u =\eta_{RL}^u=+1$ for $\Lambda=3$~TeV 
(dot-dot-dashed curve), $\Lambda=4$~TeV (long-dashed curve), and
$\Lambda=5$~TeV (dot-dashed curve).
The data points are combined H1 and ZEUS measurements.

\vskip 0.4cm

The first case is two $Z'$'s, $Z_\psi$ and $Z_\eta$ arising from the 
breaking of the grand unified group E6 \cite{er5m}.  
The best fits occur for 
$M_{Z'}< 100$~GeV which is ruled out by Tevatron data \cite{cdf}.
The second case 
is for a $Z'$ model that couples to the usual weak hypercharge (HYP)
\cite{hyp}
which has a best fit for $M_{Z'}\sim 420$~GeV.  
Although this 
particular value of $M_{Z'}$ maximized the log-likelihood function, the 
variation of $\ln {\cal L}$ with $M_{Z'}$ is so gradual that this value
should be taken as little more than suggestive 
and is only slightly more than $1\sigma$
better than the $M_{Z'}\to \infty$ case which corresponds to the 
standard model.
Because there are 
no published limits on $Z_{HYP}$ by the Tevatron collaborations
we estimated a limit using the criteria that at least 
ten dilepton events would 
be observed in the combined $e^+e^- + \mu^+ \mu^-$ 
channels in $p\bar{p}$ at $\sqrt{s}=1.8$~TeV with a total integrated 
luminosity of 110~pb$^{-1}$ \cite{godfrey}.  
We obtain the limit
$M_{Z_{HYP}}> 780$~GeV.  Although this
limit should not be taken as absolute, we have some faith in its
validity since a similar analysis gives results very close to the 
CDF limits for $Z_\psi$, $Z_\chi$, $Z_\eta$ and $Z_{LR}$.  
We conclude that the fitted value of $M_{Z_{HYP}}$ 
is also ruled out by Tevatron data.

We next ask whether our fitted $Z'$ masses are consistent with the 
energy scales extracted from the contact interaction fits \cite{ci,ci2}.  
The standard parametrization of contact interactions is 
$4\pi\eta_{ij}/\Lambda^2$ 
where $\Lambda$ is the scale of new physics and $\eta_{ij}$ denotes 
the chirality of the contact interaction.  A very massive $Z'$ 
arising from a gauge theory contributes a contact term that goes 
roughly like $e^2/(c_w^2 s_w^2 M_{Z'}^2)$.  The actual expression 
would depend on the specific model with its predicted ratio of $Z'$ to 
$Z$ couplings and fermion-$Z'$ couplings.  Rearranging this expression 
so that it is in the same form as that of the contact interaction and 
taking $M_{Z'}=420$~GeV we obtain $\Lambda\sim 2$~TeV in reasonable 
agreement with the value of $\Lambda=3$~TeV
obtained by direct fits to the contact 
interaction \cite{ci}.  The agreement would be even better if we had not 
neglected $Q^2$ dependent 
propagator effects which should be included for such 
a low mass exchange object.  For comparison purposes we also show in
Fig. 1 the 
$e^+p$ cross section with a contact interaction with $\Lambda=3$~TeV 
and $\eta_{LR}^u=\eta_{RL}^u=+1$, the value obtained by a number of 
authors \cite{ci} \footnote{We found that this contact term 
has a lower $\chi^2$ than the standard model but that
the log likelihood function indicates a worse fit.  When redoing the 
fit to $\Lambda$ with $\eta_{LR}^u=\eta_{RL}^u=+1$ we found that using 
the $\chi^2$ criteria we obtained a best fit for $\Lambda\sim 4$~TeV 
and using the log-likelihood criteria $\Lambda\sim 5$~TeV.  We would 
not regard this latter result to be statistically distinguishable from 
the SM fit. We also 
show in Fig. 1 the cross section for $\Lambda =4$ and 5~TeV.}.  
The  choice of the $\eta$'s is 
contrained by atomic parity violation measurements in cesium.
We conclude from this exercise that the 
HERA data is not compatible with a $Z'$ consistent with Tevatron 
limits, with the size of couplings 
expected in a gauge theory, but could only arise from a $Z'$ 
with strong couplings although even in this case, the statistical 
evidence is not compelling.

Composite models of gauge bosons satisfy this criteria of strong 
coupling strength.  Two such models of $Z'$'s have 
appeared in the literature; excited weak vector bosons \cite{baur} and 
the breaking electroweak symmetry strongly model (BESS) \cite{bess}.
Neither model appears to be ruled out by existing data \cite{layssac,bess2}.

To conclude,  we studied the effects of extra gauge bosons on the high 
$Q^2$ region of the $e^+p \to e^+ X$ cross section at 
$\sqrt{s}=300$~GeV.  We found that the only model 
with electroweak strength coupling typical of extended gauge theories
that gives an improved fit over the standard model fit
is ruled out by existing data from the Tevatron.  From general 
scaling arguments using the allowed contact interactions, the only 
allowed models with $Z'$'s are those with strong couplings.  Two such 
models, the excited weak boson model \cite{baur,layssac} 
and the BESS model \cite{bess,bess2} do not appear to be ruled out by 
current data.

\acknowledgments

This research was supported in part by the Natural Sciences and Engineering 
Research Council of Canada.  The author thanks Kingman Cheung for 
helpful communications.


\begin{references}

\bibitem{h197}
H1 Collaboration, C. Adloff {\it et al.}, DESY Report 97-24 (Feb. 
1997), hep-ex/9702012.

\bibitem{zeus97}
ZEUS Collaboration, J. Breitweg, {\it et al.}, DESY Report 97-25 (Feb. 
1997), hep-ex/970215.

\bibitem{lq}
J. Bl\"umlein, [hep-ph/9703287];
J. Kalinowski, R. R\"uckl, H. Spiesberger, and P.M. Zerwas, [hep-ph/9703288];
D. Choudhury and S. Raychaudhuri, [hep-ph/9703276];
J.L. Hewett and T.G. Rizzo, [hep-ph/9703337];
G.K. Leontaris and J.D. Vergados, [hep-ph/9703338];
Z. Kunszt and W.J. Stirling, [hep-ph/9703427];
I. Montvay, [hep-ph/9704280];

\bibitem{squarks}
G. Altarelli {\it et al.}, [hep-ph/9703276];
H. Dreiner and P. Morawitz, [hep-ph/9703279];
J. Kalinowski, R. R\"uckl, H. Spiesberger, and P.M. Zerwas, [hep-ph/9703288];
D. Choudhury and S. Raychaudhuri, [hep-ph/9703369];

\bibitem{ci}
V. Barger, K. Cheung, K. Hagiwara, and D. Zeppenfeld, [hep-ph/9703311];
N. Di Bartolomeo and M. Fabbrichesi, [hep-ph/9703375].

\bibitem{ci2}
K.S. Babu, C. Kolda, J. March-Russell, and F. Wilczek, [hep-ph/9703299];
M.C. Gonzalez-Garcia and S.F. Novaes, [hep-ph/9703346];
G. Altarelli {\it et al.}, [hep-ph/9703276].

\bibitem{misc}
B.A. Arbuzov, [hep-ph/9703460];
K. Akama, K. Katsuura, and H. Terazawa, [hep-ph/9704327].

\bibitem{wm}
W. Marciano,  talk given at the Pheno'97 International Symposium,
Madison WI, March 17-19, 1997.

\bibitem{c-g} For a recent review of $Z'$ physics see 
M. Cvetic and S. Godfrey, to be published in {\sl Electroweak Symmetry 
Breaking and Physics Beyond the Standard Model}, eds. T. Barklow, S. 
Dawson, H. Haber, and J. Seigrist (World Scientific, 1996) 
[hep-ph/9504216].

\bibitem{godfrey}
S. Godfrey, Phys. Rev. {\bf D51}, 1402 (1995).

\bibitem{tom}
J.L. Hewett and T.G. Rizzo, Phys. Rev. {\bf D45}, 161 (1992) gives a 
good summary of the couplings of the various models.

\bibitem{er5m}
For a detailed review see
J.L. Hewett and T.G. Rizzo, Phys. Rep. {\bf 183}, 193 (1989) and 
references therein.

\bibitem{hyp}
K.T. Mahanthappa and P.K. Mohapatra, Phys. Rev. {\bf D42}, 1732 
(1990); {\bf 42}, 2400 (1990).

\bibitem{baur}
U. Baur {\it et al.}, Phys. Rev. {\bf D35}, 297 (1987);
M. Kuroda {\it et al.}, Nucl. Phys. {\bf B261}, 432 (1985);

\bibitem{bess}
R. Casalbuoni {it et al.}, Phys. Lett. {\bf B155}, 95 (1985);
Nucl. Phys. {\bf B310}, 181 (1988);
M. Bilenky, J.-L. Kneur, F.M. Renard, and D. Schildknecht, Phys. Lett. 
{\bf B316}, 345 (1993).

\bibitem{othermodels} 
R.N. Mohapatra, 
{\sl Unification and Supersymmetry} (Springer, New York, 1986);
H. Georgi, E. Jenkins, and E.H. Simmons, Phys. Rev. Lett. {\bf 62} 
2789 (1989); V. Barger and T.G. Rizzo, Phys. Rev. {\bf D41} 956 (1990);
D. Chang, R. Mohapatra, and M. Parida, Phys. Rev. {\bf D30}, 1052 (1984);
R. Foot and O. Hern\`andez, Phys. Rev. {\bf D41}, 946 (1990);
R. Foot, O. Hern\'andez, and T.G. Rizzo, Phys. Lett. {\bf B246}, 183 (1990);
E. Ma, Phys. Rev. {\bf D36}, 274 (1987); 
K.S. Babu {\it et al.}, Phys. Rev. {\bf D36}, 878 (1987);
J.F. Gunion {\it et al.} Int. J. Mod. Phys. {\bf A2}, 118 (1987);
T. G. Rizzo, Phys. Lett. {\bf B206} 133 (1988);
A. Bagneid, T.K. Kuo, and N. Nakagawa, Int. J. Mod. Phys. {\bf A2} 
1327 (1987); {\bf 2}, 1351 (1987).

\bibitem{layssac}
J. Layssac, F.M. Renard, and C. Verzegnassi, Phys. Lett. {\bf B287}, 
267 (1992); Z. Phys. {\bf C53}, 97 (1991).

\bibitem{bess2}
R. Casalbuoni {it et al.} [hep-ph/9702325].

\bibitem{capstick}
S. Capstick and S. Godfrey, Phys. Rev. {\bf D35}, 3351 (1987);
Phys. Rev. {\bf D37}, 2466 (1988).

\bibitem{cdf}
K. Maeshima, Fermilab-Conf-96/412-E.

\end{references}
\end{document}